\documentclass[aps, twocolumn, a4paper, superscriptaddress]{revtex4-1}

\usepackage{graphicx}
\usepackage{float}
\usepackage{amsmath, amssymb}
\usepackage{lmodern}
\usepackage[T1]{fontenc}
\usepackage[utf8]{inputenc}
\usepackage[english]{babel}
\usepackage{color}

\begin{document}

\title{Invasion controlled pattern formation \\in a generalized multi-species predator-prey system}

\author{D. Bazeia}
\affiliation{Departamento de F\'\i sica, Universidade Federal da Para\'\i ba, 58051-970 Jo\~ao Pessoa, PB, Brazil}
\author{B.F. de Oliveira}
\affiliation{Departamento de F\'\i sica, Universidade Estadual de Maring\'a, 87020-900 Maring\'a, PR, Brazil}
\author{A. Szolnoki}
\affiliation{Institute of Technical Physics and Materials Science, Centre for Energy Research, Hungarian Academy of Sciences, P.O. Box 49, H-1525 Budapest, Hungary}

\begin{abstract}
Rock-scissors-paper game, as the simplest model of intransitive relation between competing agents, is a frequently quoted model to explain the stable diversity of competitors in the race of surviving. When increasing the number of competitors we may face a novel situation because beside the mentioned unidirectional predator-prey-like dominance a balanced or peer relation can emerge between some competitors. By utilizing this possibility in the present work we generalize a four-state predator-prey type model where we establish two groups of species labeled by even and odd numbers. In particular, we introduce different invasion probabilities between and within these groups, which results in a tunable intensity of bidirectional invasion among peer species. Our study reveals an exceptional richness of pattern formations where five quantitatively different phases are observed by varying solely the strength of the mentioned inner invasion. The related transition points can be identified with the help of appropriate order parameters based on the spatial autocorrelation decay, on the fraction of empty sites, and on the variance of the species density. Furthermore, the application of diverse, alliance-specific inner invasion rates for different groups may result in the extinction of the pair of species where this inner invasion is moderate. These observations highlight that beyond the well-known and intensively studied cyclic dominance there is an additional source of complexity of pattern formation that has not been explored earlier.
\end{abstract}

\maketitle

\section{INTRODUCTION}

To explain the diversity among competing species or states is a fundamental problem not only in biology, or ecology but also in social sciences \cite{chesson_ares00,hauert_s02,aguiar_n09}. One of the possible mechanisms that explains the stable coexistence of unequal species is the presence of intransitive relation or in other words cyclic dominance between competitors \cite{laird_e08,traulsen_jtb12}.  In game theory this relation can be well described by the so-called rock-scissors-paper game \cite{szolnoki_jrsif14}. Paper is cut by scissors, scissors are crushed by rock, and finally rock is wrapped by paper. In this way the circle ends and establishes the above described relation. In the absence of a superior competitor all the mentioned members can survive and hence diversity is preserved \cite{bazeia_epl18}. 

Interestingly, this relation is not a merely abstract model, but can be directly detected in several real-life systems \cite{kirkup_n04,kelsic_n15}, including microbes \cite{paquin_n83,kerr_n02}, social amoebas \cite{shibasaki_prsb18}, or even plant communities \cite{lankau_s07,cameron_jecol09}. Significant scientific efforts have been made in the last decade which clarified the possible consequences of different variations of the basic model \cite{szabo_pr07,wang_wx_pre11,szczesny_pre14,szolnoki_njp15,frey_pa10,szolnoki_pre16,valyi_15}. In spatially structured populations the topology of interaction graph is proved to be a decisive factor which determines whether an oscillatory state emerges or not \cite{masuda_prsb07,szabo_jpa04,masuda_jtb08}. Furthermore, the mobility of competing species is identified as an important factor to maintain diversity \cite{reichenbach_n07,bazeia_epl17,mobilia_g16,armano_srep17,avelino_pre18}, but some research groups also underline the nontrivial role of mutations \cite{mobilia_jtb10,park_c18,park_c18c,nagatani_jtb19}. Additionally, a recent work, obtained from off-lattice simulations, revealed the critical role of density on the original problem of maintaining diversity \cite{avelino_epl18}. It is worth noting that cyclic dominance can also emerge in systems where the values of payoff matrix, which characterizes the basic relation of different microscopic states or strategies, do not necessarily predict such interaction. Instead, this relation could be the result of a collective behavior due to the limited interactions with neighbors in a spatial system where effective multi-point interactions emerge \cite{szolnoki_pre10b,dobramysl_jpa18,szolnoki_njp14,gao_l_srep15b,roman_jtb16,szolnoki_epl15}.

Naturally, the number of competing species are  not necessarily limited to three, but can be extended to four, five \cite{roman_jsm12,lutz_jtb13,vukov_pre13,avelino_pla14,rulquin_pre14,intoy_jsm13} or even more species \cite{szabo_jpa05,avelino_pre14,szabo_pre08b,brown_pre17,avelino_pla17,esmaeili_pre18}. This makes the food-web more complex where the relation between two members is not restricted to a unidirectional predator-prey type, but also a balanced, or bidirectional relation can also emerge. This chance allows new kind of solutions, including alliances or associations, to emerge \cite{szabo_jpa05,szabo_pre08}. Beside the topological complexity of food-web an additional freedom is the heterogeneity of invasion rates between species. In some cases the latter fact alone is capable to change the final state significantly \cite{perc_pre07b,masuda_jtb08,he_q_pre10,szolnoki_srep16b,cazaubiel_jtb17,liu_a_epl17}. 

In this work we follow this research avenue and generalize a previously introduced four-species model where every species has two preys in a cyclic manner \cite{avelino_pre12b}. As a result, some relations between species become unbiased or balanced because these peer species mutually invade each other. This fact allows us to distinguish the strengths of unidirectional and bidirectional invasions and establish a tunable parameter that characterizes the inner relations of peer species.
Our key observation is the stationary pattern of the resulting evolutionary process can be varied intensively by tuning the inner invasion rate of peer species exclusively. The resulting phases can be distinguished quantitatively with the help of appropriate order parameters. These observations emphasize that not only the complex topology of a food-web, but also the varying invasion rates between related species can be the source of diverse patterns of the stationary states. 

\section{THE MODEL}

In the following we generalize a previously introduced cyclically dominated May-Leonard-type model \cite{frey_pa10} of four species \cite{avelino_pre12b}.  Initially, empty sites, labeled by 0, and all competing species, labeled by $i=1 \dots 4$, are distributed uniformly on a $L \times L$ square grid where periodic boundary conditions are applied. At each time step a randomly chosen active individual interacts with one of the four nearest neighbor passive sites by executing the following elementary steps.  

If the passive site is empty then the active individual reproduces by filling the empty site with probability $\mu$. When a motion step is applied then the active and passive individuals switch their positions with probability $m$. The last elementary step is the so-called predation when the active predator kills the passive prey and generates an empty site in the lattice.

\begin{figure}
\centering
\includegraphics{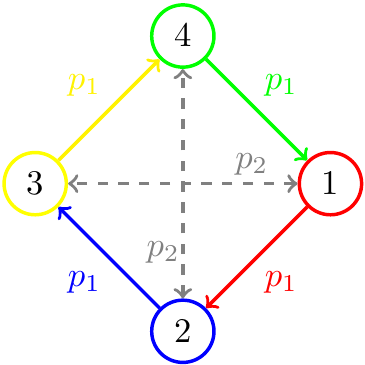}
\caption{Invasions between competing species. Solid arrows indicate the unidirectional invasions between primary predator-prey species which happen with probability $p_1$, while dashed arrows indicate bidirectional invasions between peer species that happen with probability $p_2 \leq p_1$.}
\label{def}
\end{figure}

Importantly, as an extension of the earlier introduced basic model \cite{avelino_pre12b}, we distinguish different predation probabilities between species depending on whether their labels are odd or even. In particular, as Fig.~\ref{def} illustrates, an active $i$ player predates a passive $i+1$ species and generates an empty site with probability $p_1$. However, the predation between species $i$ and species $i+2$ happens with probability $p_2$. (Naturally labels are always considered cyclically to keep $i=1 \dots 4$ interval.) 
In this way we can distinguish predation strength between predator-prey pairs where invasion is unidirectional and between peer species where bidirectional invasions can happen. The members of latter pairs, like species 1 and 3, or species 2 and 4, are equally strong because they can mutually invade each other and keep a balanced relation, as it is stressed by dashed arrows in Fig.~\ref{def}. Interestingly, such a peer pair can form a defensive alliance against an external predator species that would dominate one of the members of the mentioned pair otherwise. Just to give an example, the invasion of species 2 toward species 3 can be avoided if species 1 is present and protects peer member species 3.

Summing up our model definition, the simulation algorithm can be given as follows. At each time step an active site and a neighboring passive site are chosen randomly. After we decide whether a mobility, a reproduction, or a predation elementary step is executed. Their relative weights are: $m=0.5$, $\mu=0.25$ and $p=0.25$. If the mobility step is chosen, then the active and passive sites exchange their positions. Note that the passive site can be any individual or an empty space. If the reproduction step is chosen, then the active species can duplicate itself only if the passive site was empty. 
In case of predation step we first consider the labels of the active $i$ species and the passive neighbor. 
If the label of the passive species is $i+1$ then the latter will disappear with probability $p_1$. Alternatively, if the label of passive species is $i+2$ then it will die out with probability $p_2$. Evidently, if the passive site is occupied by a predator species of the active species, or passive and active sites are occupied by identical species, or the active site is empty then nothing happens.

In our generalized model the key parameter is the value of $p_2$, which controls the inner, or bidirectional invasion between peer species. Notably, the gradual variation of $p_2$ allows us to bridge two previously studied independent models \cite{avelino_pre12b}. More precisely, in the $p_2=0$ limit we get back the so-called $I_4$ model where partnerships of peer species, such as $\{1+3\}$ or $\{2+4\}$, emerge and occupy different spatial regions. In the other extreme limit, when $p_2=p_1=1$ the model becomes equivalent to the so-called $II_4$ model where peer domains diminish and homogeneous spirals with four-arms characterize the stationary state \cite{avelino_pre12b}. As noted, in our present work we apply a relatively high mobility rate ($m=0.5$) comparing to the basic model of Ref.\cite{avelino_pre12b}. In this way emerging spirals of invasion fronts are not suppressed by low mobility rate, as it was observed earlier.

\begin{figure}
\includegraphics[width=6.3cm]{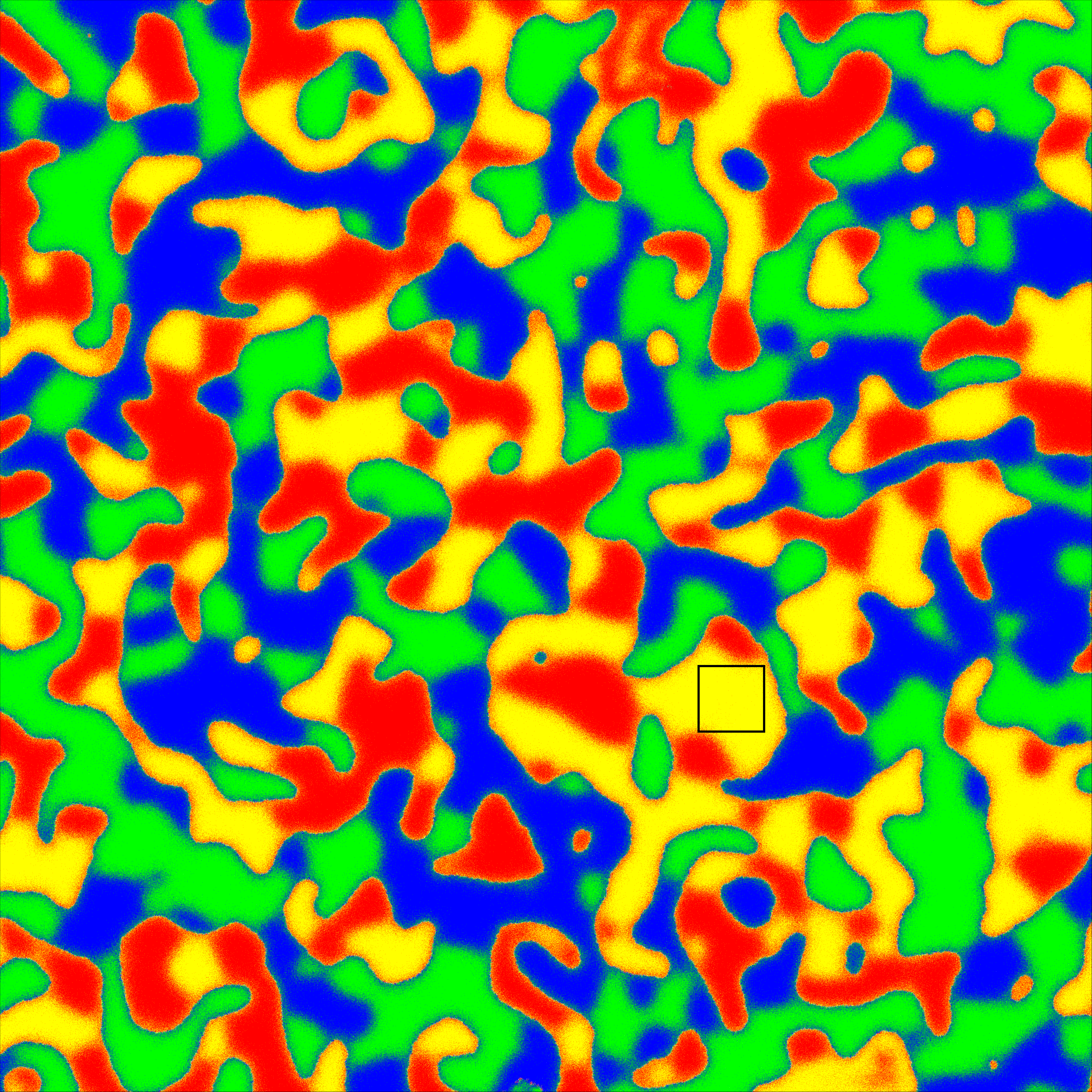}
\caption{Stationary state of an $5000 \times 5000$ system after 10000 generations obtained at $p_2=0.005$. The square in the centre of the pattern shows a $300 \times 300$ homogeneous area that is occupied exclusively by species 3. This example illustrates that the typical system size used by ordinary numerical works would result in misleading conclusions in our present model.}
\label{zoom}
\end{figure}

A full Monte Carlo step or in other words a full generation involves  $N= L \times L$ interactions or elementary steps described above. We should stress that a sufficiently high system size is necessary, otherwise we can easily obtain misleading results. To illustrate this we present the stationary pattern of an $5000 \times 5000$ system in Fig.~\ref{zoom} which was obtained at $p_2=0.005$. Here species are colored in agreement with the color-code used in model definition of Fig.~\ref{def}. The snapshot of Fig.~\ref{zoom} depicts large homogeneous domains whose linear size can easily exceed an $L=300$ lattice site (one of these spots is framed by a square of latter size). This example illustrates nicely that during the simulations we faced serious finite-size problems \cite{lutz_g17}, but luckily in the $p_2>0.01$ region $L=2000$ linear system size was generally enough to gain data, which are free from finite-size problems. 

\section{RESULTS}

\begin{figure}
\centering
\includegraphics{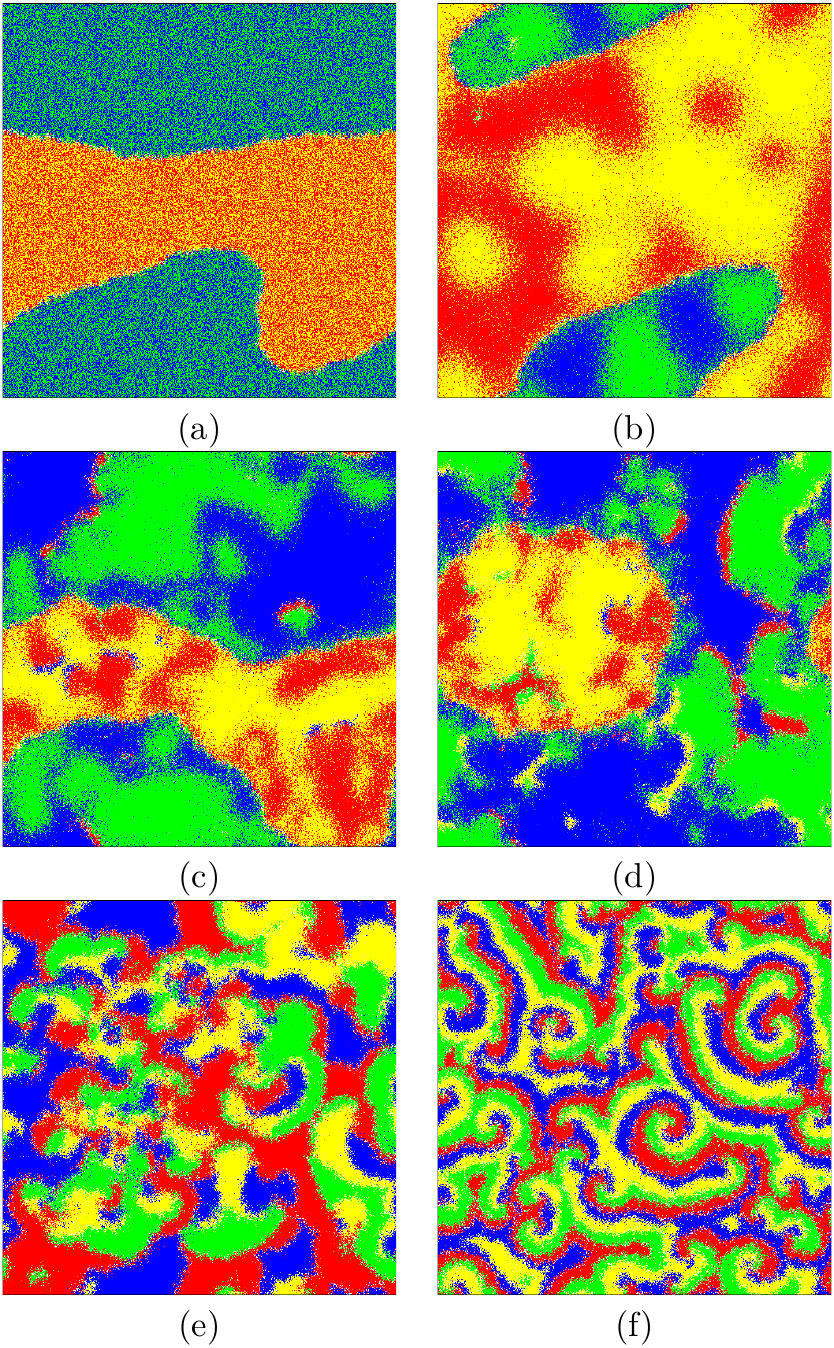}
\caption{Representative patterns of different phases in dependence of $p_2$ invasion rate. The values are $p_2=0$ (a), 0.02 (b), 0.06 (c), 0.12 (d), 0.25 (e), and 1 (f). Snapshots of stationary states were taken after 10000 generations for a $500 \times 500$ system.}
\label{snapshots}
\end{figure}

We first present our main observations how the characteristic patterns change by varying only the $p_2$ value between 0 and 1, while $p_1=1$ is kept fixed. To obtain a general overview about the emerging patterns we provide in \cite{scan} an animation showing the typical spatiotemporal patterns in dependence of $p_2$. Based on this we can identify five characteristic regions as a function of invasion strength. The typical patterns of these phases and the separated state of $p_2=0$ case are plotted in Fig.~\ref{snapshots}.

The qualitative description of different phases can be given as follows. If $p_2$ is large enough, shown in panel~(f) of Fig.~\ref{snapshots}, then we can observe clear four-color rotating spirals that characterizes typical four-state systems where species dominate cyclically each other similarly to the extended Lotka-Volterra type dynamics \cite{szabo_pre04,peltomaki_pre08,hua_epl13}.
When we start decreasing the value of $p_2$ the four-color vortices are replaced by three-color vortices, as illustrated in panel~(e) of Fig.~\ref{snapshots}. 

By decreasing the value of $p_2$ further we enter to a phase where domains composed by peer species first emerge. This phenomenon is shown in panel~(d). Since the relation of peer species is balanced therefore the borders which separate them are not as sharp as domain walls previously observed for unidirectional invasion. This effect becomes more pronounced for smaller $p_2$ values as shown in panels~(a)-(c). In parallel the three-color vortices disappear. Such vortices are always the source of propagating waves, hence in the absence of them one would expect increased characteristic length of domains. On the other hand, however, the effective mix of peer species (between 1 and 3 or between 2 and 4) is still intensive which prevents typical length from growing. Both effects are weakened if we decrease $p_2$ even further, shown in panel~(c),
which results in smooth interfaces separating domains of different peer species. Simultaneously, homogeneous spots within such a two-species domain become also larger. This state is illustrated in panel~(b) of Fig.~\ref{snapshots} signaling an enlarged typical length. Consequently, the densities of species fluctuate strongly in time which may involve serious finite size effects. For example, when the system size is comparable to the typical length of domains then the actual portions of species could be significantly different at a specific time. Such a situation is illustrated in panel~(d) of Fig.~\ref{snapshots} where the temporary portions of blue and green are seemingly higher than the portions of red and yellow colors. But we can also observe reversed effect on panel~(b) where the majority of sites are occupied by the $\{1+3\}$ alliance. Evidently, this contradicts to the basic symmetry of our model, shown in Fig.~\ref{def}, that can only be restored if the system size is large enough.

\begin{figure}
\centering
\includegraphics{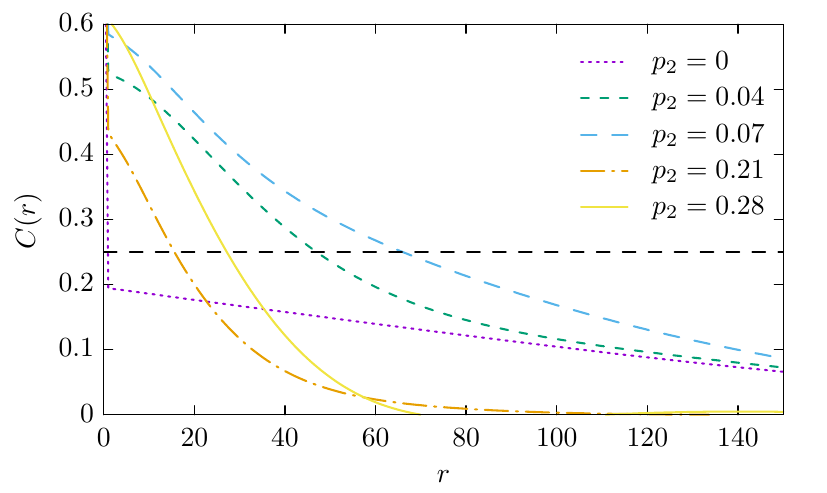}
\caption{Spatial autocorrelation functions, obtained at a $20000 \times 20000$ system size, for different values of $p_2$ when the system is in the stationary state. The dashed line, drawn at $C(r=\ell) = 0.25$, indicates the threshold value of autocorrelation which is used to define the characteristic length scale.}
\label{autocorr}
\end{figure}

As we already stressed, this enhanced characteristic length was illustrated in Fig.~\ref{zoom}. It is worth stressing that this low-$p_2$ state is significantly different from the limit case of $I_4$ model that is shown in panel~(a) of Fig.~\ref{snapshots}. In the latter case, the lack of mutual invasion between peer species results in a perfect mixture of these species, which makes the typical length fall again.

To allow readers to collect general impressions about the dynamics of pattern formation for different characteristic $p_2$ values, we provide an animation where time evolutions are shown simultaneously in \cite{multi}.

Inspired by the qualitative picture depicted above we made quantitative measurements for a more accurate description. First, we measure the typical length which characterizes the stationary states of different phases. For this goal we calculate the spatial autocorrelation function at different $p_2$ values in the long time limit when system evolved onto a stationary state. More precisely, we measure the function 

\begin{equation}
	C(r) = \displaystyle \sum_{|\vec{r}|=x+y}\dfrac{C(\vec{r})}{{\rm
min} (2N-(x+y+1), x+y+1)}\ ,
	\label{eq2}
\end{equation}
where $x$ and $y$ are the coordinates of a species in the position $\vec{r}$ on the lattice, while $C(\vec{r})$ is defined as
\begin{equation}
	C(\vec{r}) = \dfrac{1}{C(0)} \int_{\mathcal{S}}
\varphi(\vec{r})
\varphi(\vec{r}+\vec{r^{'}}) d^2\vec{r^{'}}\ .
	\label{eq1} 
\end{equation}
Here $\varphi(\vec{r}) = \phi(\vec{r}) - \langle
\phi\rangle$ and $\phi(\vec{r})$ represents the species in the position $\vec{r}$ on the lattice in the stationary state. Naturally, $\vec{r^{'}}$ spans the whole lattice, hence $\mathcal{S}$ denotes the domain of integral. Also, in agreement with general notation, $\langle \phi(t) \rangle$ represents the spatial mean value of $\phi$ 
when the system relaxed into the stationary state.
According to the model definition, we use 0 for the empty sites, and 1, 2, 3, 4 for species red, blue, yellow, and green, respectively, as also indicated in Fig.~\ref{def}.

The above defined function is plotted for some representative $p_2$ values in Fig.~\ref{autocorr}. To estimate the typical length we determine the critical $r$ value for all cases where the value of $C (r)$ function decays below the 0.25 threshold value. For comparison this value is also plotted by a horizontal dashed line in Fig.~\ref{autocorr}. As these plots illustrate, the characteristic length derived from the
autocorrelation function behaves in a largely non-monotonous way in dependence of the invasion rate $p_2$. 

\begin{figure}
\centering
\includegraphics{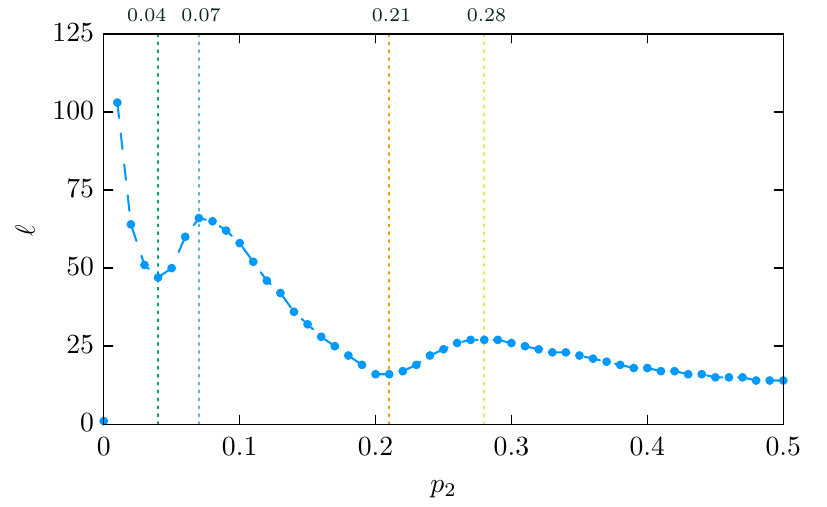}
\caption{The characteristic length, $\ell$, in dependence of control parameter $p_2$. The specific $p_2$ values where the length changes the sign of its growth tendency indicate the transition points separating different phases whose typical patterns are illustrated in Fig.~\ref{snapshots}. These values are marked by dashed vertical lines. To gain reliable results we used $20000 \times 20000$ system size. The error bars are comparable to the size of symbols.}
\label{length}
\end{figure}

This behavior becomes more transparent in Fig.~\ref{length} where the above defined characteristic length is plotted for different $p_2$ values. We note that only the $p_2<0.5$ interval is shown here because there is no observable difference between stationary states above $p_2=0.5$. In general the characteristic length decays by increasing $p_2$ value, but this curve depicts several local minimum and local maximum, which are signaled by vertical dashed lines on the plot. The related $p_2$ values are marked on the top of the figure. Importantly, these critical values mark the transition points which separate the different phases we described earlier. 

\begin{figure}
\centering
\includegraphics{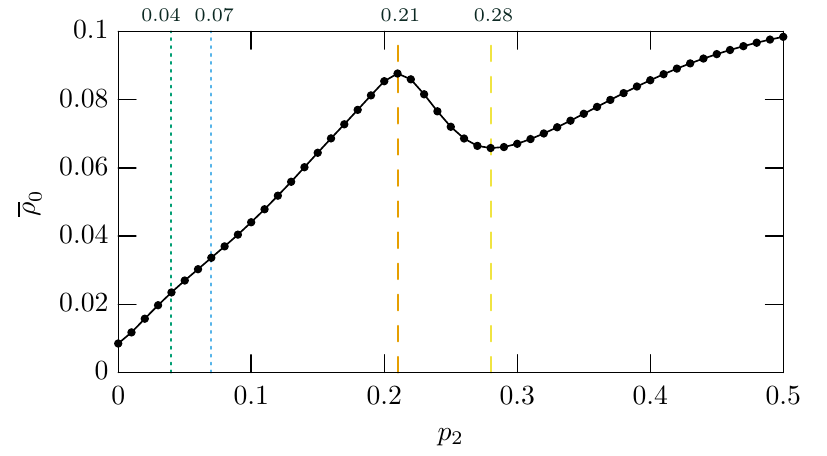}
\caption{The mean value of empty space, $\overline{\rho}_0$, as a function of $p_2$. The critical $p_2$ values where this quantity starts decaying (at $p_2=0.21$) or growing (at $p_2=0.28$) are in good agreement with the values obtained from the tendency change of characteristic length $\ell$ shown in Fig.~\ref{length}. $\overline{\rho}_0$, however, is an insensitive parameter to sign the transition points observed at small $p_2$ values. For comparison they are still marked by dotted lines in this plot.}
\label{empty}
\end{figure}

Next we also measure other parameters to confirm the importance of critical $p_2$ values we detected regarding to the characteristic length. First, we present the mean value of empty sites, $\overline{\rho}_0$, which was already proved to be an insightful quantity to characterize stationary states in previous studies \cite{avelino_pre12b,avelino_pre14}. The results for our present model are summarized in Fig.~\ref{empty}. Again, for better visibility we only show the relevant $p_2<0.5$ region here. Similarly to the characteristic length parameter, the portion of empty sites also shows a non-monotonous dependence as $p_2$ is varied. Notably, the position of the local maximum at $p_2=0.21$ and the position of the local minimum at $p_2=0.28$ are in good agreement with the critical values we found in connection to the characteristic length parameter. On the other hand, the other two critical $p_2$ values, which are also marked by vertical dotted lines in Fig.~\ref{empty}, remain hidden through the lens of $\rho_0$ parameter. The lack of observable breaking points in $\overline{\rho}_0 (p_2)$ function at small $p_2$ values suggests that when inner bidirectional invasions of peer species are too weak then the resulting concentration of empty sites becomes too small to sign the transition points reliably.

As we already argued the typical length and the resulting stationary pattern may change significantly by varying the invasion rate between peer species. This effect can be captured indirectly by measuring the standard deviation of $\rho_i(t)$ ($i \in [1,\dots,4] $) functions in the stationary state. When the typical length becomes comparable to the applied system size then the expected symmetry of four species may be broken temporarily which leads to high fluctuation in the time dependence of these functions.

\begin{figure}
\centering
\includegraphics{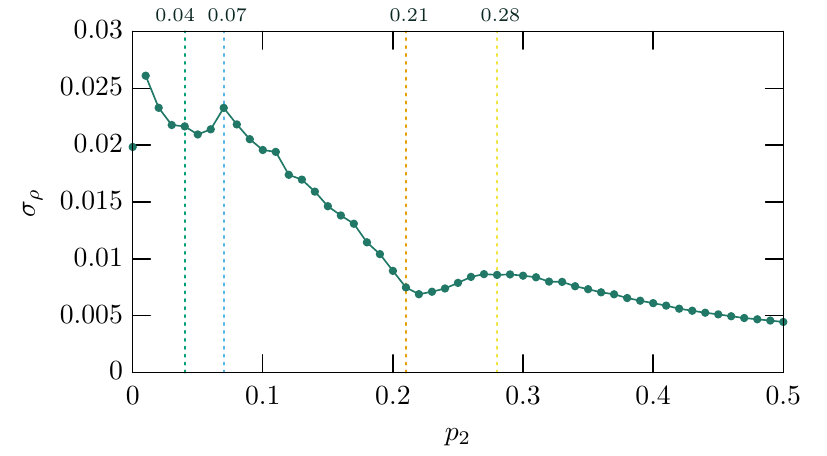}
\caption{The average of standard deviations of $\rho_i(t)$ $(i=1\dots 4)$ functions in dependence of $p_2$. Similarly to previous plots the positions of previously detected transition points are marked by vertical lines. The standard deviation is calculated from $10000$ generations in the stationary state where the curve is the average of $600$ independent runs on a $2000 \times 2000$ grid size.}
\label{fluctuations}
\end{figure}

To reveal this effect we monitored the time dependence of all $i=1 \dots 4$ species in the stationary state and calculated their standard deviations. The results for different $p_2$ values are plotted in Fig.~\ref{fluctuations}. Due to the fundamental symmetry of our model here we present only the average of standard deviations for all species, because this quantity behaves similarly for all four $i$ values. This curve basically confirms our expectation, namely, the positions of local minimum and local maximum values are in good agreement with those obtained for other quantities. 

It is worth noting that the enhanced fluctuation in the intermediate $0.04<p_2<0.21$ region is the direct consequence of how partnerships work between peer species. More precisely, as we already noted, species 1 and 3 can form a sort of alliance against species 2 and 4. If species 2 invades species 3 then a neighboring species 1 can strike back. Similarly, the invasion of species 4 against species 1 can be weakened by a neighboring species 3. If $p_2$ is small then this alliance cannot function well and the invasion fronts become smooth due to clear ranks between neighboring species. However, if $p_2$ is high enough then we get back the previously classified $II_4$ model \cite{avelino_pre12b} where homogeneous domains form four-arm spirals. Between these two extremes the partnership between peer species are functioning partly, which results in highly irregular invasion fronts and enhanced fluctuation of species. This effect can be detected clearly in Fig.~\ref{fluctuations}.

From the fact how fluctuation depends on $p_2$ and from the representative patterns of different phases shown in Fig.~\ref{snapshots} we may conclude that partnerships of peer species play a decisive role on the emergence of first two phases at small $p_2$ values. More precisely, here the expected spirals, generated by the cyclic dominance between even and odd labeled species, disappear and they are replaced by the direct competition of alliances composed by peer species. Here the yellow-red species of $\{1+3\}$ and the blue-green species of $\{2+4\}$ are equal in strength because of the symmetry of the food-web shown in Fig.~\ref{def}. This symmetry, however, can be easily broken if we apply unequal inner invasion strengths for different alliances. A conceptually similar effect has already been observed for three-member alliances in multi-species systems \cite{szabo_pre08b,szolnoki_epl15}. 
More precisely, if two cyclic dominating alliance compete then the one in which the inner invasion is faster can prevail and crowed out the alternative alliance where the inner invasion is slower \cite{perc_pre07b}.

\begin{figure}
\centering
\includegraphics{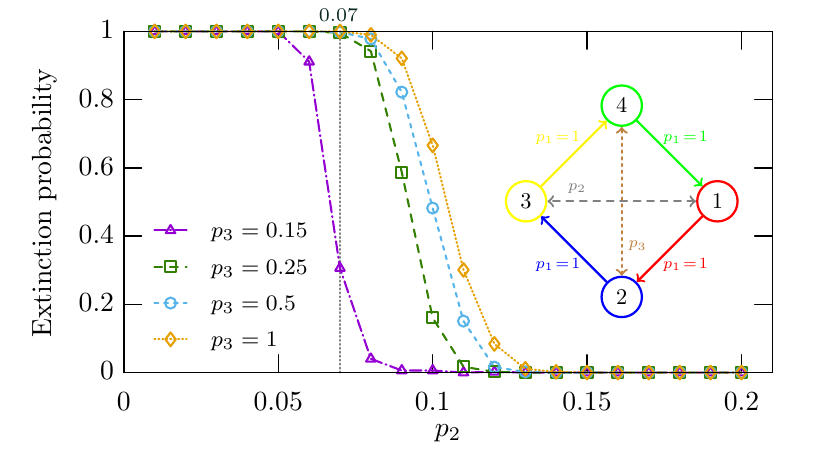}
\includegraphics{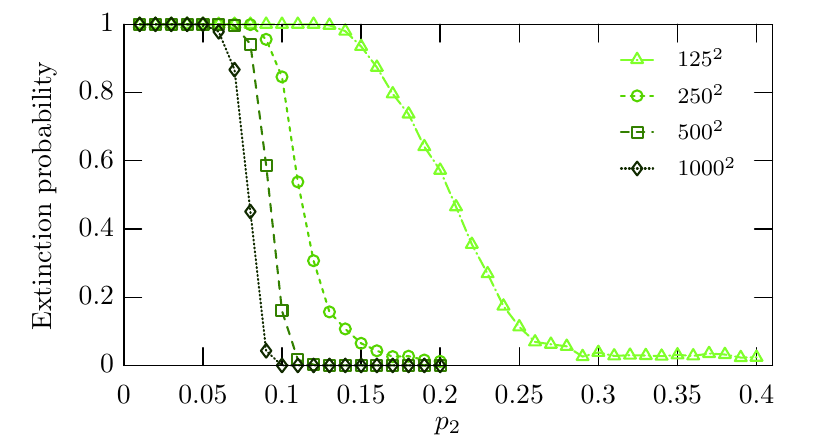}
\caption{Top panel shows the extinction probability of the $\{1+3\}$ alliance as a function of $p_2$ for different fixed values of $p_3$. In this case $p_3$ is the inner bidirectional invasion strength between species 2 and 4, while probability $p_2$ still represents the similar rate between species 1 and 3. The applied invasion rates are summarized in the inset of top panel. These results were obtained from $1000$ simulation runs for each data point in a $500 \times 500$ grid size where every simulation run is aborted after $5000$ steps if no extinction occurs. Bottom panel depicts the extinction probabilities for different system sizes at fixed $p_3=0.25$. The applied linear sizes are shown in the legend.}
\label{general}
\end{figure}

To confirm the possible conceptual similarity with our present model we generalize our model further and introduce alliance-specific inner bidirectional invasion rates in the rest of this work. In particular, we introduce $p_3 \ne p_2$ invasion rates between peer species 2 and 4 as it is shown in the inset of top panel of Fig.~\ref{general}. Technically, we keep $p_3$ constant while the value of $p_2$ is varied gradually.

As expected, the alliance of $\{1+3\}$ species cannot survive if $p_2$ is too small comparing to $p_3$ because they are dominated by the $\{2+4\}$ alliance where inner invasion, hence the resulting mix of species, is more intensive. The probability of the extinction for different fixed $p_3$ values is plotted in the top panel of Fig.~\ref{general}. Here an individual simulation was aborted after 5000 steps if no extinction occurred. The plotted values are the average of 1000 independent runs at fixed system size. We stress that the extinction of $\{1+3\}$ species is not a finite-size effect in the present case, as may happen even for the symmetric $p_2=p_3$ model if the system size is too small. Instead, in the present non-symmetric case it is a straightforward consequence of the dominance of $\{2+4\}$ alliance. Naturally, the expected extinction time may depend on the system size, but the extinction probability function converges to a limit case as we increase the system size gradually. This phenomenon is illustrated in the bottom panel of Fig.~\ref{general}, where we plotted the extinction probabilities for different $L$ values at fixed $p_3=0.25$ value. This plot demonstrates that the usage of $L=500$ linear size can predict the large system size limit qualitatively well.

As Figure~\ref{general} suggests the critical $p_2$ value where the original four-species system becomes a two-species system is decreasing as we decrease $p_3$. In the limit case it tends to $p_2 \approx 0.07$ which is the transition point between the second and third phases in the symmetric model. This behavior indirectly supports our previous conjecture that the patterns characterize the low $p_2$ value regime is principally determined by the competition of alliances composed by peer species.

\section{DISCUSSION AND CONCLUSIONS}

To maintain biological and ecological diversity is a fundamental challenge for mankind and this cannot be solved without gaining deeper insights about the basic mechanisms which drive permanent evolution. The problem is hard because interactions among competitors can easily result in a complicated food-web with subtle topology. 
For instance, closed loops in such food-webs can provide a higher level of complexity that cannot be observed in a system where the food-web is characterized by a tree-like graph. 
In particular, in the presence of loops new kind of solutions, like cyclic time development of competing species may emerge. 
But beyond topological obstacles an additional difficulty can also emerge when the intensities of interaction are significantly different among competing species or agents \cite{szabo_pre08}.

In this work we followed the latter research path by generalizing a previously established model of four interacting species with intransitive relations \cite{avelino_pre12b}. Our main motivation was to distinguish the bidirectional inner invasion rate between peer species and the unidirectional invasions characterize primary predator-prey partners. In this way the resulting mixing between peer species, who form a protective alliance against external species, can be tuned via a single parameter. 

According to our key observation the strength of inner invasion within an alliance of peer species can play a decisive role on the resulting stationary state and several quantitatively different characteristic patterns can be detected as the related control parameter is varied. In these phases the microscopic mechanisms which are responsible for the emerging pattern can be different. Rather counter-intuitively the primary unidirectional predator-prey type invasions become dominant when the mutual invasions within peer species are intensive, while the competition of alliances acts as the leading pattern formation process when this inner bidirectional invasion is moderate. 

In dependence of the mentioned $p_2$ control parameter we have observed five distinct phases where the emerging spatiotemporal patterns are different. The related transition points which separate these phases can be detected accurately by introducing appropriate order parameters. The characteristic length, which is calculated from the spatial autocorrelation function, is proved to be the most sensitive parameter which signals all emerging transition points. The measuring of standard deviation of time-dependent density functions of competing species is also proved to be an effective quantity to detect these transition points. The breaking points of latter parameter agree with those predicted by the $p_2$ dependence of characteristic length. For sake of completeness we have also measured the mean value of empty sites, which was reported as
a useful parameter to quantify stationary states in earlier studies \cite{avelino_pre14,avelino_pre12b}. The $p_2$ dependence of this quantity signals some of these transition points at the same positions as they were marked by the previously mentioned quantities. This parameter, however, becomes ineffective to sign transition points when its average value is too small due to the moderate inner invasion between peer species.  

We have generalized our model further by introducing alliance-specific inner invasion strengths, hence the resulting effective mixture between species 1 and 3 become different from the inner mixture of species 2 and 4. In this way we can break the fundamental symmetry between competing alliances and demonstrate that it has a decisive role on the final outcome if the strengths of inner invasion rates are different enough. 
Indirectly, the latter observation also supports our argument that in the low $p_2$ value region the leading mechanism which determines the pattern formation is the competition of alliances formed by peer species.

From these observations we can conclude that the diverse invasion strengths between predator-prey partners may play an important role on the final state similarly to the pure topology of food-web. Therefore the classification of stable solutions based solely on the geometry of interactions is not satisfactory and more careful investigations are necessary when we try to predict the final stable solutions of a multi-species interacting system. 

\section*{ACKNOWLEDGMENTS}

This research was supported by the Brazilian agencies CAPES, CNPq, Funda\c c\~ao Arauc\'aria, INCT-FCx, Paraiba State Research Foundation (Grant 0015/2019) and by the Hungarian National Research Fund (Grant K-120785).

\end{document}